\documentclass[twocolumn,showpacs,preprintnumbers,amsmath,amssymb,prb]{revtex4}
\usepackage[dvips]{color,graphics,epsfig,rotating}
\usepackage{graphicx}% Include figure files
\usepackage{dcolumn}% Align table columns on decimal point
\usepackage{bm}% bold math

\begin{document}

\title{Polar Behavior in a Magnetic Perovskite Via A-Site Size Disorder}

\author{D.J. Singh}
\affiliation{Materials Science and Technology Division,
Oak Ridge National
Laboratory, Oak Ridge, Tennessee 37831-6114} 

\author{Chul Hong Park}
\affiliation{Research Center for Dielectric and Advanced Matter Physics,
Pusan National University, Busan 609-735, Korea}

\date{\today} 

\begin{abstract}
We elucidate a mechanism for obtaining polar behavior in magnetic
perovskites based on A-site disorder and demonstrate this mechanism
by density functional calculations for the double perovskite
(La,Lu)MnNiO$_6$ with Lu concentrations at and
below 50\%.
We show that this material combines polar behavior and
ferromagnetism. The mechanism is quite general and may be
applicable to a wide range of magnetic perovskites.
\end{abstract}


\maketitle

There is great interest in materials that combine
magnetism and polar behavior, especially
multiferroics with both ferromagnetism and ferroelectricity.
\cite{astrov,folen,fiebig,eerenstein,hill,spaldin}
While there are many ferromagnets and many ferroelectrics, there
remarkably few materials combining the two properties.
Arguments have been made explaining this apparent incompatibility
in the case of perovskite
$AB$O$_3$ oxides.
\cite{hill}
In essence, it is because the best perovskite
magnets have magnetic ions on the $B$ site, while ferroelectrics, such
as BaTiO$_3$, usually have $B$ site ions with no $d$ electrons.
Here we elucidate a mecahanism for inducing polar behavior
in magnetic perovskites and demonstrate this mechanism by density
functional calculations for a ferromagnetic Ni-Mn double perovskite.
This mechanism is quite general and may be applicable to a wide
range of magnetic perovskites.

Perovskite lattice instabilities are often understood using the
tolerance factor $t=(r_{\rm O}+r_{A})/\sqrt{2}(r_{\rm }+r_{B})$,
where $r_{\rm O}$, $r_A$, and $r_B$ are the O, $A$-site and $B$-site
ionic radii, respectively. \cite{shannon}
Ferroelectrics, such as BaTiO$_3$
and KNbO$_3$, have $t > 1$, indicating the that $B$ site ion is too
small for its site in the ideal cubic structure. In the ferroelectric
ground state of these
so-called $B$-site driven materials, this ion off-centers, aided by
hybridization with O states.
There is another important class of ferroelectric perovskites,
so called $A$-site driven materials. In these, $t$ is normally
less than unity, and the ferroelectricity is from off-centering
of $A$-site ions. This family includes the technologically important
Pb based piezoelectrics and relaxor ferroelectrics. 
The essential physics is lone pair stereochemistry,
specifically hybridization of Pb and Bi 6$p$ states with O $p$ states.
\cite{cohen}
This class includes the few known magnetic ferroelectrics
with strong ferroelectric properties, {\em e.g.} BiFeO$_3$, BiMnO$_3$,
and PbVO$_3$.
\cite{smith,smolenskii,chiba,shpanchenko,belik}
Without Pb or Bi, $t<1$ perovskite structures generally
derive from $B$O$_6$ octahedral tilts and not $A$-site off-centering.

Ions with $d$ electrons are generally larger than $d^0$ ions.
The majority of magnetic perovskites have $t<1$,
with lattice structures based on tilts of the $B$O$_6$ octahedra
and not ferroelectricity.
However, first principles calculations have shown that, while these
materials have tilted ground states, if the octahedra are prevented
from tilting, strong ferroelectricity may
result, with an energy intermediate between
the ideal cubic perovskite structure and the ground state structure, but
closer to the later. \cite{halilov,singh-05,bilc,imf}
The role of Pb and Bi is then to shift the balance
between these states to yield ferroelectricity.
Here we use a different approach to shift
the balance between these states based on $A$-site
size disorder. \cite{singh-05,bilc,imf}
This is applied rare-earth double perovskites,
$R_2$MnNiO$_6$ where we obtain
polar behavior combined with ferromagnetism for mixtures of large
and small rare earth ions.
The motivation for this choice is that the charge difference
$\delta Q$=2 between Mn$^{4+}$ and Ni$^{2+}$ and their size
difference (Shannon radii,
\cite{shannon} $r_{\rm Mn^{4+}}$=0.67\AA, 
$r_{\rm Ni^{2+}}$=0.83\AA) indicates $B$-site ordering into the double
perovskite structure,
\cite{anderson}
and that La$_2$MnNiO$_6$ and Bi$_2$MnNiO$_6$
are known to form and to be ferromagnetic.
\cite{g61,blasse,azuma,hughs,rogado,sakai}
These two compounds were previously studied by first principles calculations.
\cite{uratani,matar}

\begin{figure}[tbp]  % h=here, b=bottom, p=page
\epsfig{file=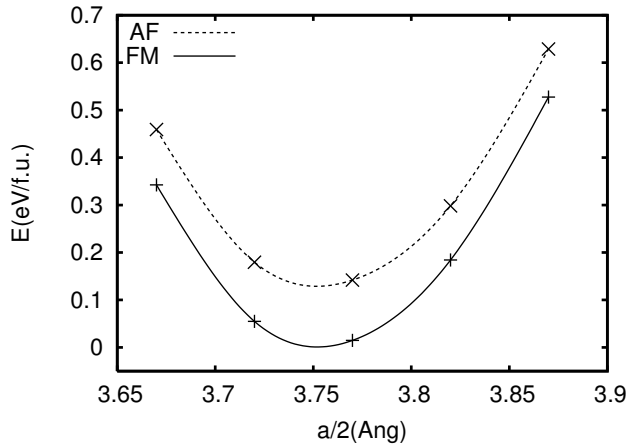,angle=-90,width=3.4in}
\caption {\label{volume}
Energy of 10 atom LaLuMnNiO$_6$ cells vs.
pseudocubic lattice parameter ($a$/2) for parallel (FM) and
antiparallel (AF) Ni and Mn moments.
Note that the FM alignment is strongly favored.
}
\end{figure}

We used the local density
approximation (LDA) in the general potential linearized augmented planewave
(LAPW)
method, \cite{singh-book}
with well converged basis sets including local orbitals.
\cite{singh-lo}
The LAPW sphere radii were
2.0 $a_0$ for La and Lu, 1.9 $a_0$ for Ni and Mn and 1.55 $a_0$ for O.
\cite{small}
Mn and Ni atoms were placed in supercells with the double
perovskite (rock-salt) ordering, and various orderings of the
$A$-site ions. The primary results reported here were done with
40 atom supercells. However, the lattice parameter was determined by
relaxation of a 10 atom cell of compositions LaLuMnNiO$_6$.
No symmetry was imposed, either for the
10 atom or 40 atom cells, 
but for the 40 atom cells
the lattice parameters were held fixed at their pseudocubic values
as determined from the relaxation of the 10 atom ferromagnetic cell,
which as shown in Fig. \ref{volume} was 3.75 \AA.
For both the parallel and antiparallel spin alignments, relaxation
of the 10 atom cells yielded structures with off-centering of
the Lu ions, {\em i.e.} polar structures, even though
perovskite $R_{25}$ type tilts are allowed in this cell.

\begin{figure}[tbp]  % h=here, b=bottom, p=page
\epsfig{file=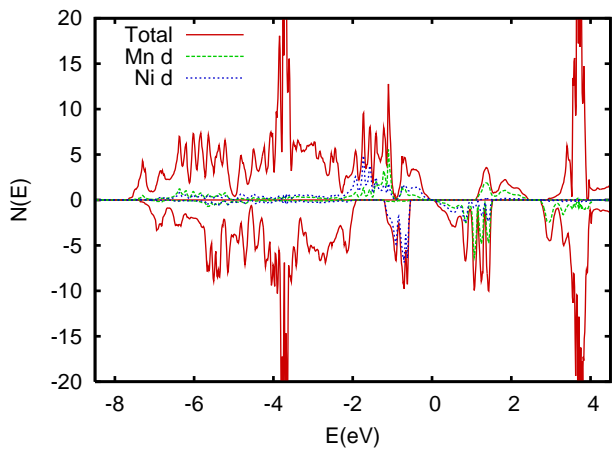,angle=-90,width=3.4in}
\caption {\label{lda-dos} (color online)
Electronic density of states of
a 10 atom ferromagnetic LaLuMnNiO$_6$ cell ($a$=3.77\AA).
The peaks at $\sim$ -3.7 eV and $\sim$ 3.7 eV are
the Lu and La $f$ states, respectively.
The projections are onto LAPW spheres.
}
\end{figure}

\begin{figure}[tbp]  % h=here, b=bottom, p=page
\epsfig{file=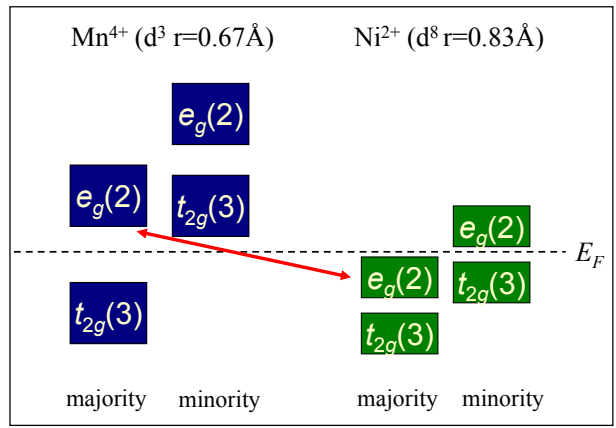,angle=0,width=3.2in}
\caption {\label{exch} (color online)
Schematic depiction of the electronic structure of LaLuMnNiO$_6$
showing the dominant superexchange coupling by the arrow.
}
\end{figure}

Our electronic structure for (La,Lu)MnNiO$_6$
is similar to those previously found for the Bi and La
analogues,
\cite{uratani,matar}
and show high spin Mn$^{4+}$ and Ni$^{2+}$.
The LDA density of states for the relaxed ferromagnetic structure
at $a$=3.77\AA~ is shown in Fig. \ref{lda-dos}, and schematically
in Fig. \ref{exch}.
For the various supercells we find either
very small gaps or small band overlaps in the LDA depending on
the exact crystal structure. We also performed
some LDA+U calculations (not shown). These yield insulating band
structures, with gaps depending on $U$.
Near metallicity is highly unfavorable for ferroelectricity, as it
means strong electronic dielectric screening that will weaken the Coulomb
interactions. Nonetheless, we use the LDA to avoid
the ambiguity associated with an adjustable
parameter ($U$) and expect that the
prediction of polar behavior will be robust.

\begin{figure}[tbp]  % h=here, b=bottom, p=page
\epsfig{file=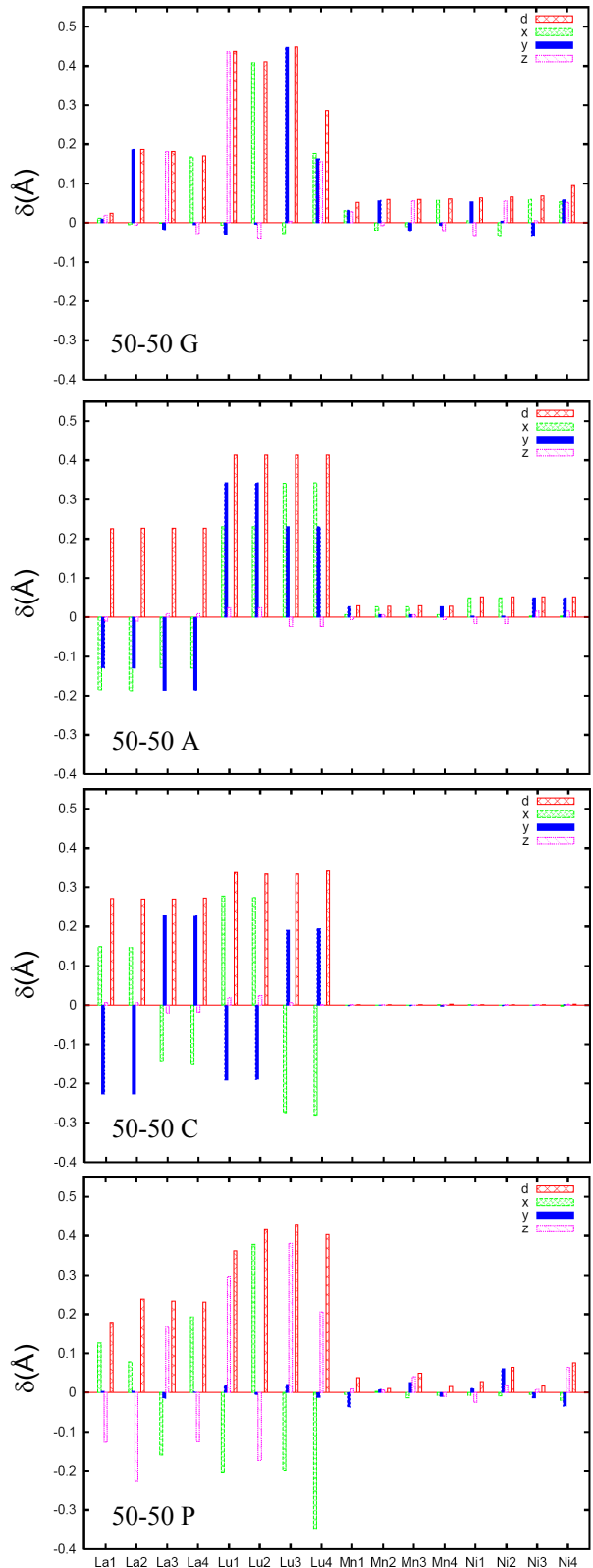,angle=0,width=3.2in}
\caption {\label{disp-q} (color online)
Cation displacements with respect
to their O cages in a 40 atom relaxed supercells of composition
(La$_{0.5}$Lu$_{0.5}$)MnNiO$_6$.
}
\end{figure}

The ferromagnetism is due to the fact that with parallel alignments
of the Mn and Ni spins there is a strong cross-gap hybridization
of the unoccupied $e_g$ majority states of Mn$^{4+}$ with the
occupied majority $e_g$ states of Ni$^{2+}$,
leading to a ferromagnetic coupling consistent also with
the Goodenough-Kanamori rules.
\cite{good,kan,good-book}
This ferromagnetic superexchange is particularly strong because
in perovskites the strongest coupling through the near
linear $B$-O-$B$ bonds is via $e_g$ - $p_\sigma$ hopping.
The strength of
this coupling is evident from the sizable crystal field splittings
\cite{cf-note} of both
the Mn and Ni $d$ states (Fig. \ref{lda-dos}).
Thus the hybridization between occupied and unoccupied $e_g$ orbitals,
allowed for ferromagnetic alignment, but not for antiferromagnetic alignment,
strongly favors ferromagnetism. Since the
hopping is mediated by O this is not direct exchange, like the weak
ferromagnetic coupling generally associated with the Goodenough-Kanamori
ferromagnetism, but is a conventional strong superexchange.
\cite{o-note,good-book,anderson-se}

% \begin{figure}[tbp]  % h=here, b=bottom, p=page
% \epsfig{file=disp-t.eps,angle=0,width=3.2in}
% \caption {\label{disp-t} (color online)
% Cation displacements with respect
% to their O cages in a 40 atom relaxed supercell of composition
% (La$_{5/8}$Lu$_{3/8}$)MnNiO$_6$.
% }
% \end{figure}

% \begin{figure}[tbp]  % h=here, b=bottom, p=page
% \epsfig{file=disp-d.eps,angle=0,width=3.1in}
% \caption {\label{disp-d} (color online)
% Cation displacements with respect
% to their O cages in a 40 atom relaxed supercell of composition
% (La$_{0.75}$Lu$_{0.25}$)MnNiO$_6$.
% The displacements are shown along the Cartesian directions, and
% in addition the magnitude of the displacement (d) is given.
% }
% \end{figure}

\begin{table}[tbp]
\caption{Lu displacements for the various supercells. $|<\delta>|$ is
the magnitude of the average Lu displacement, and $<|\delta|>$ is the
average magnitude.}
\label{tab-disp}
\vspace{0.15in}
\begin{tabular}{lccccc}
\hline
$n_{\rm La}$ & $n_{\rm Lu}$ & &
 ~$|<\delta>|$~~~ & ~~$<|\delta|>$~ & $|<\delta>|/<|\delta|>$~ \\
\hline
4 & 4 & ``G" & 0.243\AA & 0.396\AA & 0.61 \\
4 & 4 & ``A" & 0.405\AA & 0.414\AA & 0.98 \\
4 & 4 & ``C" & 0.013\AA & 0.396\AA & 0.04 \\
4 & 4 & ``P" & 0.201\AA & 0.403\AA & 0.48 \\
5 & 3 &      & 0.379\AA & 0.379\AA & 1.00 \\
6 & 2 &      & 0.331\AA & 0.331\AA & 1.00 \\
\hline
\end{tabular}
\end{table}

While our relaxed structure for the 10 atom
cell is polar, this size supercell favors ferroelectricity
(ferroelectricity arises from a zone center instability,
while the competing tilt modes occur at the zone boundary and a 10 atom
cell restricts tilts to the $R$-point).
We therefore performed
supercell calculations with a $2\times 2\times 2$ 40 atom supercell.
This cell is doubled along the [001], [011] and [111] directions
and so allows arbitrary mixtures of tilt instabilities and accommodates
the observed Glazer patterns of perovskite tilt systems.
\cite{glazer}

As mentioned, all supercells were for the double perovskite structure,
{\em i.e.} rock-salt ordering of the $B$-site Ni and Mn
ions. For the A-site, we considered
four arrangements of the Lu and La at a 50--50 composition as well
as one supercell each for 5/8--3/8 and 3/4--1/4 compositions.
The specific cells at the 50--50 composition were 
(1) rock-salt ordering of La and Lu
(``G" in the following),
(2) (001) layers of La and Lu (``A"),
(3) lines of La and Lu along [001] and ordering $c(2\times 2)$ in plane
(``C")
and 
(4) a cell maximizing like near neighbors
(La at corner and edge centers, Lu at face and body centers, denoted ``P").
The 5/8--3/8 supercell was constructed by replacing one Lu by La in the
cell ``G", while the 75--25 supercell was made by substitution of two
Lu by La in the same ``G" supercell.

Fig. \ref{disp-q}
shows the cation positions
in the lowest energy relaxed structures for the 50\% Lu supercells
with respect to
the centers of their nearest neighbor O cage (nearest 12 O atoms for the
$A$-site atoms, and nearest 6 O for the $B$-sites ions).
The average displacements of the Lu for the various cells are
sumarized in Table \ref{tab-disp}.
As may be seen,
in all cases Lu strongly off-centers, by on average
$\sim$ 0.4\AA~ for the 50--50
supercells and slightly less for the lower concentration cells.
Interestingly, unlike other $A$-site driven perovskite ferroelectrics,
\cite{ghita}
there is very little off-centering of the $B$-site ions.
In fact the largest off-centering aside from Lu are of the La ions
and depending on the supercell these may or may not be parallel to the
Lu.

The individual Lu off-centerings tend to avoid
[111] and equivalent directions, and
also tend to
be non-collinear with each other for 50\% Lu concentration, e.g.
in the ``G"
ordering (nominally the highest symmetry case), three
of the Lu displace along different Cartesian directions, while the
fourth has a smaller displacement near [111].
A preference for Cartesian
directions was noted in
(K,Li)NbO$_3$, \cite{bilc} and understood from
the fact that the square faces of the cage (the faces
with the most room for the Li ion) are along these directions.
Turning to the question of polar behavior, both of the cells at
less than 50\% Lu concentration show polar structures. At
50\% the cells ``G", ``P" and ``A" have polar structures,
most strongly so for ``G", while ``C" is nearly antiferrodistortive, with
tilts that avoid compressing the O - La bonds.
Of the four supercells investigated the relaxed ``G" structure had the highest
energy. Taking this energy as the zero, the calulated energies were
-0.43 eV, -0.36 eV, and -0.42 eV, for ``A", ``C" and ``P", repectively
on a per formula unit (10 atom) basis.
The similarity of the energies other than ``G" show that the $A$-site
ions will be disordered in material made by conventional methods, in
agreement with what would normally be expected in perovskites with chemically
similar, same charge, ions.
While in the ordered ``C" structure
structure a long range tilt pattern of this type is allowed, this
will not be the case in general for disordered $A$-site ions.

Assuming that the $A$-sites are disordered in the alloy, this
dependence of polar behavior on chemical ordering
is more consistent with relaxor ferroelectric behavior than
normal ferroelectricity. \cite{smol54,burns2,samara03}
However, it should be kept in mind that the relaxations reported
here were performed within the LDA, which also predicts zero or very
small band gaps, while
in reality larger, but unknown gaps may be present. Larger gaps
would lower the electronic dielectric constant favoring ferroelectricity
and stronger coupling between the Lu off-centerings. While actual
ferroelectricity may occur, what can be concluded here
is that polar behavior will occur in
disordered (La,Lu)MnNiO$_6$ for Lu concentrations at or below 50\%.
This may be ferroelectricity or relaxor ferroelectricity.

This polar behavior arises because of frustration of the
tilt instabilities due to the mixture of $A$-site cation sizes
and the fact that
the coherence length for off-centering of $A$-site ions is shorter
than that for the tilt instabilities.
This mechanism is quite general in principle,
may be useful in producing polar behavior
in other perovskites.
Qualitatively, this is related
to the rigidity of the $B$O$_6$ octahedra. This condition is often
but not always met, as for example, while often tilt instabilities are
strengthened by pressure, there are cases where this does not hold.
\cite{angel}

It may be difficult to synthesize perovskite (La,Lu)$_2$MnNiO$_6$
due to phase separation \cite{park}
or competing phases, {\em e.g.}
tungsten bronze as often occurs with mismatched $A$-sites.
These issues can sometimes be overcome using thin film techniques,
such as pulse laser deposition, or by high pressure synthesis, which
favors the high density perovskite structure. Further, polar behavior
is predicted over a wide composition range. This may help in finding
specific compositions amenable to synthesis.
In any case, the proposed mechanism is quite general, and
should apply to other mixtures of $A$-site ions with different
size, {\em e.g.} La with other small rare earths.

This work was supported by the Department of Energy,
Division of Materials Science and Engineering and the Office
of Naval Research.


\begin{references}
\bibitem{small}
Smaller radii of 1.53$a_0$ for O and 1.88$a_0$ for the $B$-site
ions were used in some relaxations to avoid overlapping spheres.
The results for different radii were cross-checked.
\bibitem{o-note}
Note the exchange splitting of O $p$ bands (Fig. \ref{lda-dos}),
which shows the involvement of O via hybridization.
\end{references}
\end{document}